# High-Throughput Studies of Novel Magnetic Materials in Borides

Zhen Zhang[1], Kirill D. Belashchenko[2], Vladimir Antropov[1,3]

[1]Department of Physics and Astronomy, Iowa State University, Ames, IA 50011, USA
[2]Department of Physics and Astronomy and Nebraska Center for Materials and Nanoscience, University of Nebraska-Lincoln, Lincoln, NE 68588, USA
[3]Ames National Laboratory, U.S. Department of Energy, Ames, IA 50011, USA

Borides are a versatile material family with various properties for valuable applications. Conventional magnetism, such as ferromagnetism and antiferromagnetism in borides, have been extensively studied. However, research on unconventional magnetism in borides where quantum effects are dominant is scarce. Here, we implement a high-throughput workflow combining first-principles calculations, materials prediction, and magnetic properties calculations to discover novel magnetism and magnetic materials in borides. Successfully applying the workflow, we report three families of novel magnetic borides, including two families of borides exhibiting quantum magnetism. One is a family of dimerized quantum magnets among $YCrB_4$-type borides, which provides a rare platform for studying the spin-gap quantum critical point. The other is a family of altermagnets among $FeMo_2B_2$-type borides, extending the magnetic orderings exhibited by borides beyond conventional ferromagnetism and antiferromagnetism. We also predict a family of magnetic laminate transition metal borides, known as the MAB phases, in the $AlFe_2B_2$-type family, which provide pure-phase or alloying candidates for studying magnetocaloric materials and the associated magnetic transitions. The workflow is expected to be used in further studies of novel magnetism and magnetic materials.

## I. Introduction

Borides exhibit a variety of structures, chemical compositions, and properties. One of their most important applications is in permanent magnets, such as $Nd_2Fe_{14}B$, which remains one of the best permanent magnets to date. Due to the significance of permanent magnets, borides containing rare earth and 3d transition metals and conventional magnetism, such as ferromagnetism, antiferromagnetism, ferrimagnetism, and paramagnetism, have been extensively studied. However, there has been little research into the unconventional magnetism of borides. For this reason, we conduct high-throughput studies on novel magnetism and magnetic materials in borides using first-principles calculations.

Here, we report three families of novel magnetic borides: (1) We unveil a family of dimerized quantum magnets in the $YCrB_4$-type structure [1]. This finding enables research in borides for spin-gap quantum critical point (QCP) and potential Bose-Einstein condensation (BEC) of magnetic excitations. (2) We discuss a family of altermagnets in the $FeMo_2B_2$-type structure. This finding enables research in borides for altermagnetism and potential next-generation spintronics and magnonics. (3) We predict a family of magnetic MAB (laminate transition metal boride) phases in the $AlFe_2B_2$-type structure. This finding enables research in borides for magnetocaloric effect and potential 2D magnetic materials.

## II. Methodology

To effectively and efficiently conduct the high-throughput search for novel magnetism and magnetic materials, we develop the following step-by-step workflow and apply it to various families of materials. First, we select structural prototypes with interesting structural features that may lead to unusual magnetic phenomena. This will be demonstrated later in this work. Second, we perform high-throughput elemental substitutions in the selected structural prototypes and evaluate the thermodynamic stability of each new material by computing the relative formation energy distance to the known ternary formation enthalpy convex hull. Known convex hull phases are obtained from materials databases such as the Materials Project [2] and the Open Quantum Materials Database (OQMD) [3]. The formation energies of all phases are recalculated by spin-polarized density functional theory (DFT) using the same DFT settings. If the formation energy is below the known convex hull, the new material is assumed to be stable. In this case, the convex hull is reconstructed to include the new material's formation energy, the distance of which from the convex hull, also known as stability, is now zero. If the formation energy is within 50 meV/atom above the convex hull, the material is considered to be metastable and potentially synthesizable. If the formation energy is within 200 meV/atom above the convex hull, the material is considered potentially metastable and worth reporting. If the formation energy is more than 200 meV/atom above the convex hull, the material is deemed unstable and excluded from further consideration. Nevertheless, some borides predicted to be unstable have been synthesized in experiments [4,5]. Third, for all stable and metastable materials, we carry out high-throughput collinear magnetic calculations for various magnetic configurations. Magnetic configurations are carefully constructed to include all combinations of parallel and antiparallel spins between up to at least the third nearest neighbors of the magnetically active elements. Then, the magnetic ground state is determined based on the lowest self-consistent total energy. Fourth, for the magnetic ground state, we conduct detailed examinations of its physical properties. These examinations include the magnetic moments, electronic structure, magnonic structure, magnetic exchange coupling, and symmetry analysis. This way, novel magnetism and novel magnetic materials suitable for synthesis are predicted.

DFT calculations are performed using the projector augmented wave (PAW) method as implemented in the VASP package. The electronic exchange-correlation energy is treated by the Perdew-Burke-Ernzerhof (PBE) generalized gradient approximation (GGA). The convergence thresholds are set to be $10^{-5}$ eV for electronic self-consistency and 0.01 eV Å$^{-1}$ for ionic relaxation. The kinetic energy cutoff for the plane-wave-basis set and the k-point grid for the Brillouin zone sampling are set to be sufficiently high and dense, respectively, to achieve the convergence of energy within 1 meV/atom.

### III. RESULTS AND DISCUSSION

#### A. A Family of Dimerized Quantum Magnets in the YCrB$_4$-Type Structure

Quantum magnets are of great interest due to their many exotic phenomena. One of the simplest kinds of quantum magnets is those consisting of strongly antiferromagnetically coupled spin dimers, known as dimerized quantum magnets. Dimerized quantum magnets close to the QCP between a disordered singlet spin-dimer phase, with a spin gap, and the ordered conventional Néel antiferromagnetic (AFM) phase are of great significance, which provide a unique platform to study BEC of magnetic excitations [6]. So far, dimerized quantum magnets have been discovered mostly in oxides and halides [6], such as BaCuSi$_2$O$_6$, Ba$_3$Mn$_2$O$_8$, and TlCuCl$_3$. To the best of our knowledge, dimerized quantum magnets have not been found among borides before.

Here, we unveil a family of dimerized quantum magnets in the YCrB$_4$-type borides [1]. YCrB$_4$ and related materials of this structural type have been known for decades [7,8]. However, investigations of the electronic and magnetic properties of these compounds are scarce, even for the prototypical YCrB$_4$. A remarkable structural feature of this class of materials is a transition-metal dimer. Whether such a dimer can be magnetized and even form a quantum spin dimer remains unknown and is yet to be explored. Motivated by this idea, we carry out high-throughput DFT calculations to search for stable and metastable YCrB$_4$-type compounds, MTB$_4$ (M = Sc, Y, La, Ce, Lu, Mg, Ca, Al; T = V, Cr, Mn, Fe, Co, Ni), with antiferromagnetically coupled 3d atoms within each dimer as the ground state.

Fig. 1(a) shows the two ground-state magnetic configurations with antiferromagnetically coupled atoms within the dimer. They are labeled as AFF and AAA, respectively, according to our convention. AFF does not have alternating spin orientations along the out-of-plane direction, while AAA does. The arrangements of the in-plane spin dimers are also different between AFF and AAA.

Fig. 1(b) shows the phase stabilities of the nine dimerized quantum magnets found in MTB$_4$. Four compounds are stable: ScCrB$_4$, YCrB$_4$, LuCrB$_4$, and MgMnB$_4$. Their formation energies are part of their respective convex hull, and thus, their stabilities are denoted as zero. Five compounds are metastable: MgCrB$_4$, CaCrB$_4$, AlCrB$_4$, LaCrB$_4$, and CaMnB$_4$. YCrB$_4$ [9] is known to exist, and LuCrB$_4$ [10] was also briefly mentioned as existing. All of them are potentially synthesizable. Fig. 1(c) shows the magnetic moments on the 3d transition metal atoms. The magnetic moments are all about 1 $\mu_B$, suggesting the spin-1/2 state of the T atom in these magnetic compounds. The ground state is AAA for MgCrB$_4$ and CaCrB$_4$ and AFF for the other seven compounds. This class of compounds obeys a simple isoelectronic rule to form dimerized quantum magnets. M = IIIA and IIIB elements, Al, Sc, Y, La, and Lu, form dimerized quantum magnets only with Cr. M = IIA elements, Mg and Ca, form dimerized quantum magnets with both Cr and Mn.

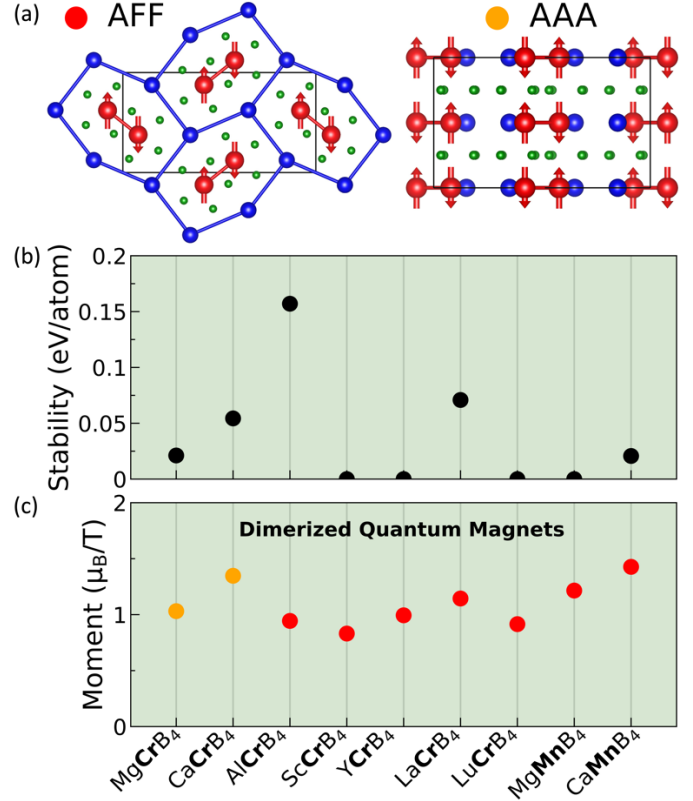

Fig. 1. (a) DFT ground-state magnetic configurations of the YCrB$_4$-type structure and the associated symbols and labels [1]. Blue, red, and green spheres represent the Y, Cr, and B sites, respectively. (b) Phase stability of YCrB$_4$-type MTB$_4$. (c) Magnetic moment on the magnetically active 3d transition metals in the DFT ground state.

With this isoelectronic rule in mind, we select three representative compounds, YCrB$_4$, MgCrB$_4$, and MgMnB$_4$, to compute their magnetic exchange coupling parameters in the DFT ground state. Exchange parameters are calculated using the linear response technique [11] implemented within the tight-binding linear muffin-tin orbital (TB-LMTO) method in the Questaal code [12]. A strong AFM coupling within the dimer promotes the formation of a singlet spin-gap state, resulting in the absence of magnetic order [13,14]. The singlet spin-gap state is in competition with the antiferromagnetically ordered state, which is supported by the interdimer exchange coupling. Following the approach in [14], we examine a spin-1/2 Heisenberg lattice Hamiltonian, where interactions between dimers are treated at the mean-field level, while the 4 × 4 Hamiltonian of each individual dimer is exactly diagonalized. In this dimer Hamiltonian, interactions with other dimers are approximated by effective fields, where the spin operators for neighboring dimers are replaced by their expectation values. A QCP between the singlet spin-gap state and the magnetically ordered state is predicted by the solution to this quantum mean-field model at $|J'_0/J_D| = 1$, where $J_D$ is the intradimer exchange coupling, and $J'_0$ is the "non-staggered" total exchange coupling of a given magnetic moment to the rest of the lattice excluding its intradimer neighbor. YCrB$_4$, MgCrB$_4$, and MgMnB$_4$ have $|J'_0/J_D|$ = 1.06, 1.10, 0.83. Thus, they are expected to be close to the QCP.

By alloying this family of dimerized quantum magnets, it may be possible to tune the exchange coupling across the QCP. This family of dimerized quantum magnets provides a rare platform for studying the spin-gap QCP.

### B. A Family of Altermagnets in the FeMo$_2$B$_2$-Type Structure

Recently, a new class of collinear magnetism, known as altermagnetism, has been discovered [15,16]. This form of magnetism exhibits momentum-dependent electronic band spin splitting, a hallmark of ferromagnetism, while maintaining zero net magnetization, characteristic of antiferromagnetism. Altermagnets can exhibit spin-polarized currents and anomalous Hall effect without a large net magnetization. Their distinct magnetic and electronic properties hold great promise for next-generation spintronic devices. So far, altermagnets have been found mainly in compounds containing group VA and VIA elements.

Here, we describe a family of altermagnets with the FeMo$_2$B$_2$-type structure. FeMo$_2$B$_2$ and related materials of this structural type have been known for decades [17,18]. Calculations have been conducted to study some fundamental physical properties, including magnetism, of some materials of this family [19]. Altermagnetism has been theoretically found in FeNb$_2$B$_2$ and FeTa$_2$B$_2$ [20,21]. No recent synthesis nor measurement for the predicted magnetism is available, either. We perform high-throughput DFT calculations to seek stable and metastable FeMo$_2$B$_2$-type compounds, TM$_2$B$_2$ (T = V, Cr, Mn, Fe, Co, Ni; M = Y, Zr, Nb, Mo, Tc, Ru, Rh, Pd, Ag, Cd, La, Hf, Ta, W, Re, Os, Ir, Pt, Au, Hg), with ground-state magnetic orderings classified as altermagnetism.

Fig. 2(a) shows the ground-state magnetic ordering for a number of compounds in this family. This ordering has ferromagnetically coupled T chains along the c-axis and AFM interactions between the nearest chains. The magnetically inactive M atoms at the Mo site and the magnetically active T atoms at the Fe site form tetragonal spin sublattices. Opposite-spin sublattices in real space are connected by rotation or mirror symmetries, which is characteristic of altermagnets. Correspondingly, the opposite-spin electronic states in momentum space are also connected by rotation or mirror symmetries. Fig. 2(b) shows the spin-polarized nonrelativistic electronic band structures of FeNb$_2$B$_2$. The spin splitting can reach as large as 0.2 eV near the Fermi level.

Fig. 2(c) shows the phase stabilities of eleven altermagnets found in the TM$_2$B$_2$ structure. MnMo$_2$B$_2$ is stable. The rest ten compounds, MnRu$_2$B$_2$, MnW$_2$B$_2$, MnRe$_2$B$_2$, FeZr$_2$B$_2$, FeNb$_2$B$_2$, FeMo$_2$B$_2$, FeRu$_2$B$_2$, FeHf$_2$B$_2$, FeTa$_2$B$_2$, and FeW$_2$B$_2$ are metastable. All of them are potentially synthesizable. Fig. 2(d) shows the magnetic moments on the 3d transition metals. Except for FeZr$_2$B$_2$ and FeHf$_2$B$_2$, these compounds have rather large magnetic moments of about 2 $\mu_B$.

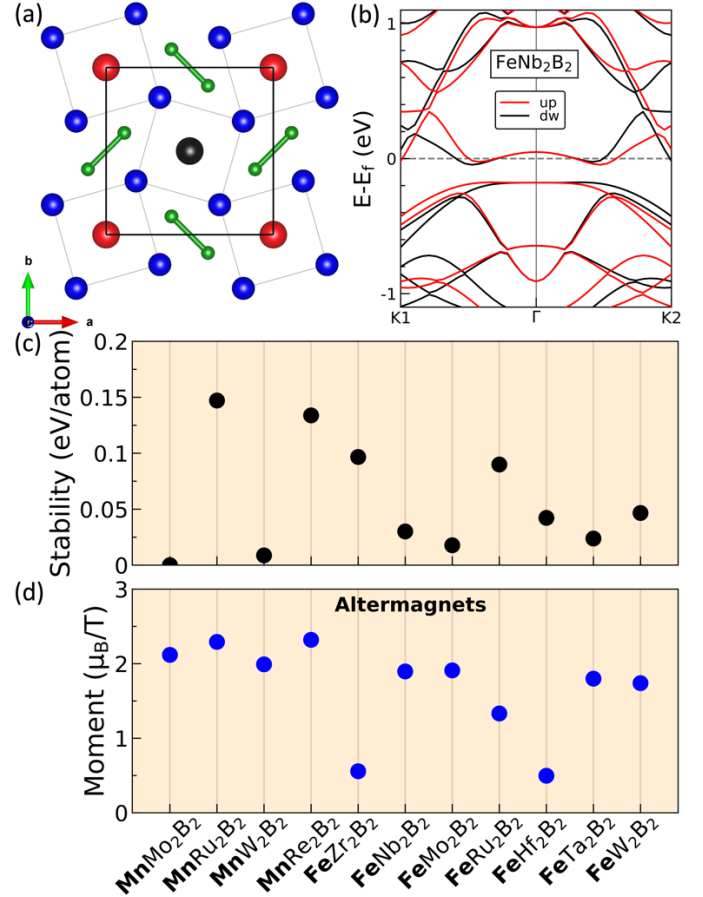

Fig. 2. (a) DFT ground-state magnetic configuration of the FeMo$_2$B$_2$-type structure. Red and black spheres represent the Fe sites with opposite spins. Blue and green spheres represent the Mo and B sites, respectively. (b) Spin-polarized electronic band structure of the DFT ground-state FeNb$_2$B$_2$. Red and black curves show bands of opposite spins. K-points in the Brillouin zone: Γ(0, 0, 0), K1(3/8, 1/4, 0), and K2(-3/8, 1/4, 0). (c) Phase stability of FeMo$_2$B$_2$-type TM$_2$B$_2$. (d) Magnetic moment on the magnetically active 3d transition metals in the DFT ground state.

This family of altermagnets provides a rare platform for studying altermagnetism, chiral magnons, and potential spintronics applications in borides.

### C. A Family of Magnetic MAB Phases in the AlFe$_2$B$_2$-Type Structure

2D carbides and nitrides, known as MXenes, show promising mechanical, electrical, and magnetic properties. They have been found to be used in a wide range of applications, such as energy storage and harvesting, catalysis, electromagnetic interference shielding and antennas, optoelectronics, sensors, and medicine [22]. MXenes can be separated from the laminate MAX phases with the chemical formula M$_{n+1}$AX$_n$, where M is transition metals, A is group-A elements, and X is carbon or nitrogen. In MAX phases, due to much stronger M-X bonds than A-X and A-M bonds, the MX layer tends to be separated from the A layer. The structural features of MAX inspire the proposal and discovery of MAB phases, where B is boron instead of carbon or nitrogen [23]. Since then, people have spared no effort in finding new MAB phases and separating the boron analogs of MXenes, known as MBenes, from the MAB phases [24].

MAB phases generally have the formula $(MB)_2A_y(MB_2)_x$ [23]. One of the simplest kinds of MAB phases is $AlFe_2B_2$-type borides. Recently, $AlFe_2B_2$ has been identified as a strong magnetocaloric material comprising earth-abundant elements with a paramagnetic to ferromagnetic (FM) transition at 307 K [25]. This motivates the exploration of other compounds of this class of materials as potential strong magnetocaloric materials or alloying endmembers with $AlFe_2B_2$. Moreover, if MBenes can be successfully exfoliated from this class of magnetic MAB phases, then these magnetic MBenes could exhibit unusual 2D magnetic properties.

Here, we predict a family of magnetic MAB phases in the $AlFe_2B_2$-type borides. We conduct high-throughput DFT calculations to search for stable and metastable $AlFe_2B_2$-type compounds, $AM_2B_2$ (A = Li, Na, K, Rb, Cs, Be, Mg, Ca, Sr, Ba, Al, Si; M = V, Cr, Mn, Fe, Co, Ni), with a magnetic ground state.

Fig. 3(a) shows the $AlFe_2B_2$-type crystal structure and all the ground-state magnetic configurations of the stable and metastable compounds found in this family. In this type of structure, the A atoms at the Al site form an A layer sandwiched by two MB layers at the FeB sites. The main difference between the $AlFe_2B_2$-type MAB phase and typical MAX phases is that the A-B interactions in the $AlFe_2B_2$-type structure are significant, which prevents easy exfoliation to obtain 2D layers.

Fig. 3(b) shows the low-energy $AM_2B_2$ phases found. Three of them are stable: $AlMn_2B_2$, $BeFe_2B_2$, and $AlFe_2B_2$. $AlMn_2B_2$ [26] and $AlFe_2B_2$ [25] have been synthesized in experiments. Six compounds are metastable: $AlV_2B_2$, $LiMn_2B_2$, $SiMn_2B_2$, $LiFe_2B_2$, $MgFe_2B_2$, $SiFe_2B_2$.

Fig. 3(c) shows the magnetic moments on the 3d transition metals, M. Moments of this class of magnetic MAB phases range from 0.6 to 1.4 $\mu_B$. $AlFe_2B_2$ is FM as the ground state, which agrees with measurements [25]. $AlMn_2B_2$ is AFM as the ground state, which also agrees with some of the latest measurements [26]. The specific ground-state magnetic orderings for $AlMn_2B_2$, $SiMn_2B_2$, and $SiFe_2B_2$ are also in agreement with previous calculations [27].

This family of magnetic MAB phases provides abundant candidates for exploring novel magnetocaloric materials and alloying effects with the known $AlFe_2B_2$-type phases. If exfoliation is successful, these materials could also be used to study 2D magnetic materials.

## IV. CONCLUSION

In summary, to explore boride systems' novel magnetism and magnetic materials, we develop a workflow encompassing crystal structure analysis, high-throughput thermodynamic stability calculations, high-throughput magnetic structure calculations, and magnetic ground state analysis. Utilizing the workflow, we study three families of novel magnetic borides: dimerized quantum magnets in the $YCrB_4$-type family, altermagnets in the $FeMo_2B_2$-type family, and magnetic MAB phases in the $AlFe_2B_2$-type family. These findings potentially enable new research in borides in areas such as spin-gap QCP, BEC of magnetic excitations, altermagnetism, next-generation spintronics, magnonics, magnetocaloric effect, and 2D magnetism. With high efficiency, the developed workflow is expected to discover more novel magnetism and magnetic materials in material families, including but not limited to borides.

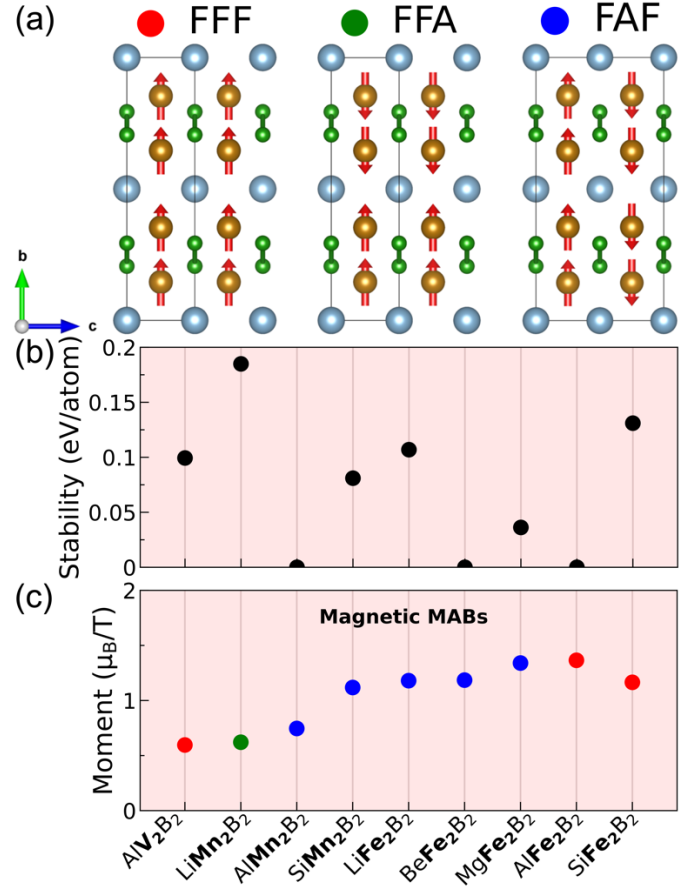

Fig. 3. (a) DFT ground-state magnetic configurations of the $AlFe_2B_2$-type structure and the associated symbols and labels. Light blue, brown, and green spheres represent the Al, Fe, and B sites, respectively. (b) Phase stability of $AlFe_2B_2$-type $AM_2B_2$. (c) Magnetic moment on the magnetically active 3d transition metals in the DFT ground state.

ACKNOWLEDGMENT

This work is supported by the U.S. Department of Energy (DOE) Established Program to Stimulate Competitive Research (EPSCoR) Grant No. DE-SC0024284. Computations were performed at the High Performance Computing facility at Iowa State University and the Holland Computing Center at the University of Nebraska. VA is supported by the U.S. DOE,

Office of Basic Energy Sciences, Division of Materials Sciences and Engineering. Ames National Laboratory is operated for the U.S. DOE by Iowa State University under Contract No. DE-AC02-07CH11358.

DATA AVAILABILITY

The data that supports the findings of this study are available within the article.